# Synergistic Role of Electron and Photon Dose in Stepwise Laser-Induced Complete Deoxygenation of Graphene Oxide Revealed by In-situ TEM

Israt Ali[1], Kenneth R. Beyerlein[1]*

[1]Institut National de la Recherche Scientifique (INRS), Center Énergie Matériaux Télécommunications, Varennes, Québec, J3X 1P7, Canada

*Corresponding Author: kenneth.beyerlein@inrs.ca

**Abstract**

Laser-induced reduction of graphene oxide (GO) represents a highly promising route to graphene synthesis, offering spatially localized processing, elimination of hazardous chemical reagents, and compatibility with ambient conditions. Here, we introduce a stepwise laser reduction strategy employing a 532 nm pulsed laser, monitored in real-time by in situ dynamic transmission electron microscopy (DTEM). By systematically varying the pulse sequence and the cumulative photon and electron dose, complete deoxygenation of GO is achieved while preserving film integrity. Core-loss EELS confirms full removal of oxygen functional groups and restoration of the $sp^2$ graphitic network, evidenced by a $\pi^*$–$\sigma^*$ energy separation of 7.0 eV, in close agreement with graphite (7.1 eV). Crucially, the cumulative electron dose is identified as an active parameter governing the reduction mechanism: electron beam exposure accounts for approximately 5 at. % of the initial oxygen removal and synergistically lowers the energy barrier for subsequent laser-driven deoxygenation, while excessive electron exposure compromises film integrity through crack formation. The optimal configuration achieves complete deoxygenation at a cumulative photon dose of $2.5 \times 10^3$ mJ/cm² with superior in-plane crystallographic order and minimal beam-induced thinning. This work establishes a versatile multi-parameter strategy for controlled scalable graphene synthesis via combined electron beam and laser irradiation.

## 1. Introduction

Transforming graphene oxide (GO) to reduced graphene oxide (rGO) represents a critical step in unlocking the full potential of graphene-based materials for electronic and optoelectronic applications[1, 2]. GO, typically synthesized via the chemical exfoliation of graphite, possesses a rich and tunable surface chemistry arising from a high density of oxygen-containing functional groups, including epoxide, ether, and hydroxyl (−OH) moieties distributed across the basal plane, and carboxyl (−COOH) and carbonyl (C=O) functionalities predominantly decorating the sheet edges[3-7]. While these oxygen functionalities confer significant chemical versatility, they simultaneously introduce a high density of structural defects that disrupt the conjugated $sp^2$ network, rendering GO electrically insulating and severely curtailing its utility in device-oriented applications. Restoring the graphitic framework through controlled deoxygenation is therefore essential to recover the exceptional electrical conductivity and carrier mobility intrinsic to pristine graphene.

Considerable efforts have been devoted to developing effective reduction strategies; however, each conventional approach carries significant limitations[8]. Chemical reduction using hydrazine or sodium borohydride, while capable of selectively removing certain oxygen functionalities, raises serious concerns regarding toxicity and environmental hazard[9, 10]. Thermal reduction methods, although widely employed, typically demand temperatures exceeding 1000 °C and involve complex processing protocols, yet frequently yield only partially reduced products characterized by substantial residual disorder, a consequence of rapid, uncontrolled CO and $CO_2$ evolution during high-temperature treatment[11-13]. These inherent shortcomings collectively underscore the pressing need for a fundamentally different reduction paradigm: one that enables precise, controlled deoxygenation and progressive graphitic reconstruction without reliance on hazardous chemical reagents, extensive thermal annealing, or elaborate preprocessing steps.

Photoreduction has emerged as an attractive alternative for the efficient conversion of GO into rGO[14]. Unlike conventional chemical or thermal approaches, laser-based reduction offers a unique combination of spatial precision and process versatility: the ability to raster the laser beam and modulate the illumination pattern enables direct writing of conductive rGO circuits within insulating GO films, opening pathways toward maskless, lithography-free device fabrication[15-17]. In addition, laser-induced reduction does not require solvents or harsh chemical reagents, making the process consistent with the principles of green chemistry. A wide range of laser wavelengths, spanning from ultraviolet (UV) to near-infrared (NIR), and pulse durations including continuous wave, nanosecond, picosecond, and femtosecond regimes, have been explored for GO reduction[18]. For example, Maren et al.[17] employed femtosecond laser pulses (280 fs) at 1064 nm to investigate GO reduction, while Chii-Rong et al.[19] demonstrated the formation of highly reduced GO using a high-repetition-rate (100 kHz) UV laser at 355 nm. Similarly, De lima et al.[20] reported that both infrared (1064 nm) and visible (532 nm) irradiation promote $sp^3$-to-$sp^2$ conversion primarily through photothermal processes, whereas UV irradiation enhances oxygen removal through a stronger photochemical contribution. Despite these advances, the complete removal of oxygen functionalities from GO remains challenging. Most reported approaches leave a residual amount of oxygen even after extensive irradiation, regardless of the laser wavelength, pulse duration, or intrinsic structure of GO. These limitations highlight the need for a new photoreduction strategy capable of driving the reduction process to completion while simultaneously enhancing the localized $sp^2$ character, ultimately enabling the fabrication of higher-quality graphene structures.

Further, GO reduction is commonly characterized using techniques such as X-ray photoelectron spectroscopy (XPS), X-ray diffraction (XRD), Raman spectroscopy, and thermogravimetric analysis (TGA) [21-23]. While these methods provide valuable information on overall chemical composition and structural evolution, they typically yield spatially averaged data and cannot resolve local variations in oxidation within heterogeneous GO films. In contrast, transmission electron microscopy (TEM) combined with electron energy loss spectroscopy (EELS) offers nanoscale insight into both structure and bonding, enabling direct evaluation of local oxidation states. For example, Anna et al. showed that fine structure features in the C K-edge can distinguish different oxidation states in GO and rGO[24]. Furthermore, in situ TEM-EELS allows dynamic monitoring of reduction processes; Mario et al. used this approach to study thermally induced GO reduction up to 1200 °C and identified temperature-dependent removal of oxygen functional groups[25].

In our previous work, laser-induced reduction of GO was investigated using multiple laser pulses at 532 nm, monitored in situ by transmission electron microscopy. By varying laser fluence between 6.36 and 15.88 mJ/cm², we quantified the removal of distinct oxygen functional groups via EELS, with estimated film temperatures ranging from 342°C to 823°C. While the process demonstrated precise control over chemical composition and followed double-exponential reduction kinetics, complete removal of all oxygen functional groups could not be achieved without compromising the structural integrity of the film[26]. In addition, the effect of the electron beam on GO and the deoxygenation of GO was not well understood. It is still unclear how oxygen removal progresses locally under controlled laser irradiation and whether the electron beam used for EELS characterization influences the reduction pathway. This raises two key questions: How does the photoreduction of GO evolve at the nanoscale, and to what extent does electron beam exposure contribute to or modify the observed reduction kinetics? Addressing these questions is essential for developing controlled photoreduction strategies capable of achieving complete deoxygenation while preserving the structural integrity of the graphene lattice.

In this work, we present a coupled electron beam and laser irradiation strategy for photoreduction that enables the complete removal of oxygen from GO while promoting the formation of localized $sp^2$-bonded carbon domains. We employ a dynamic transmission electron microscope (DTEM) integrated with EELS, which allows laser irradiation and nanoscale chemical analysis to be performed within the same instrument. The DTEM platform enables direct observation of irreversible transformations such as photoreduction, transient structural ordering, and rapid crystallization under controlled laser exposure. In our approach, nanosecond laser pulses at a wavelength of 532 nm are focused onto the GO specimen, and a novel strategy for laser irradiation is employed where the pulse fluence is increased progressively in a stepwise manner, rather than applied at a constant high value. This controlled delivery of energy enables the gradual removal of oxygen functional groups while minimizing ablation and preserving the structural integrity of the sheets. By systematically varying the cumulative photon dose through different pulse sequences, we further demonstrate that achieving complete deoxygenation at a reduced cumulative photon dose is most effectively accomplished with lower pulse sequences. Simultaneously we show that minimizing the cumulative electron dose enhances the localized $sp^2$ character of the reduced film. Finally, TEM imaging, selected area electron diffraction (SAED), and EELS collectively confirm the structural and chemical quality of the rGO produced using this strategy. These results establish a controlled pathway toward fully reduced graphene structures and highlight the importance of stepwise laser irradiation strategy for achieving complete and localized restoration of graphitic bonding.

## 2. Experimental Section

The graphene oxide was purchased from Sigma Aldrich (Item: 777676, 4mg/ml). A 100-μl volume of GO from the stock solution was dissolved in 1 ml of water to dilute it. Further, a 10-μl sample was drop cast on a lacey carbon TEM grid to avoid any other carbon contamination and dried overnight in a closed environment. Finally, the sample was placed in the TEM to study the laser-induced reduction of GO.

To track the progressive removal of oxygen and to elucidate the synergistic interplay between cumulative electron dose and cumulative photon dose, in situ DTEM combined with EELS was employed to monitor the chemical evolution of GO in real time. The DTEM at INRS is based on a Jeol JEM-2100 Plus that has been modified by incorporating mirrors into the column to allow for illumination of the sample with a pump laser, as well as the possibility of operation in pulsed photoemission mode, a schematic of which is shown in our previous article [26]. In this study, the microscope was operated in continuous beam, thermionic mode, with an accelerating voltage of 200 kV. The pump laser system was based on a nanosecond pulsed neodymium-doped yttrium-aluminum garnet (Nd:YAG) laser (Northrup Grumman). The fundamental wavelength (1064 nm) was frequency doubled to 532 nm using a Type-I barium borate (BBO) crystal. The laser was operated at a repetition rate of 10 Hz and produced a maximum second-harmonic power of 3 mW. The duration of the second harmonic pulse was measured to be 11 ns (full-width at half-maximum) with a photodiode. The laser focus on the sample has been measured to have a diameter ($1/e^2$) of 180 μm, determined by characterization of a virtual focus and critical pulse energies necessary to melt aluminum and gold films. In our experiment, the exposure of the sample was controlled by a shutter system that selected individual pulses from the laser, while the laser fluence on the sample was controlled by attenuating the emitted laser beam using a half-wave plate to rotate the polarization with respect to a thin film polarizer.

Electron energy loss spectra were measured using a Gatan Quantum GIF system. During spectra collection, an objective aperture was inserted corresponding to a semi-collection angle of 12.7 mrad. Literature values for the inelastic mean free path ($\lambda$) in graphitic materials vary from 112 nm to 120 nm, depending on the material and technique employed [27-31]. We then assumed a value of $\lambda$ = 115 nm, which was determined from measurements of graphitic films up to 200 layers, and falls within this range[30]. The oxygen concentration was quantified from core-loss EELS spectra by integrating the area under the O K-edge after background subtraction.

We tested the following different sample irradiation parameters for the complete removal of oxygen and higher localized $sp^2$ reconstruction. 1) Laser pulse sequence - defined as the number of consecutive laser pulses delivered to the GO film at a fixed fluence prior to each EELS acquisition. The number of pulses per sequence determines the incremental photon dose deposited at each reduction step. 2) Irradiated laser dose – defined as the product of the number of laser pulses delivered (N) and the fluence per pulse ($\Phi_p$). This quantity serves as a measure of the total laser energy incident on the sample driving the photoreduction process.  3) Cumulative electron dose - defined as the total number of electrons per unit area ($e^-/nm^2$) incident on the GO film. This electron beam irradiation is accumulated as a direct consequence of repeated EELS spectrum acquisitions during the reduction process. It serves as a quantitative measure of the total electron beam exposure experienced by the sample and is governed by the frequency of EELS acquisition throughout the experiment. A comprehensive summary of all experimental parameters, including pulse sequence, cumulative photon dose, number of EELS spectra collected, and cumulative electron dose for each measurement scheme tested in this study is provided in **Table 1**.

**Table 1.** Summary of irradiation parameters for all experimental schemes under high and low electron dose conditions. H and L denote high and low cumulative electron dose conditions, respectively, while the number denotes the number of laser pulses in a sequence. The table includes the total laser-irradiated dose required for complete deoxygenation, the number of EELS spectra collected, and the resulting cumulative electron dose deposited in the GO film.

| Scheme | Pulse sequence | Fluence steps ($mJ/cm^2$) | Irradiated laser dose [$N\Phi_p$] ($mJ/cm^2$) | Number of EELS spectra collected | Cumulative electron dose ($e^-/nm^2$) |
|---|---|---|---|---|---|
| H5 | 5 | 3.93<br>7.86<br>11.79<br>15.72<br>19.65 | 2240 | 43 | $3.2\times10^5$ |

| H10 | 10 | 3.93<br>7.86<br>11.79<br>15.72<br>19.65 | 3654 | 37 | $2.8×10^5$ |
|---|---|---|---|---|---|
| H20 | 20 | 3.93<br>7.86<br>11.79<br>15.72<br>19.65<br>23.58<br>27.51 | $1.4×10^4$ | 53 | $3.9×10^5$ |
| H50 | 50 | 3.93<br>7.86<br>11.79<br>15.72<br>19.65<br>23.58 | $2.9×10^4$ | 49 | $3.6×10^5$ |
| L5 | 5 | 3.93<br>7.86<br>11.79<br>15.72<br>19.65 | 2240 | 6 | $4.5×10^4$ |
| L50 | 50 | 3.93<br>7.86<br>11.79<br>15.72<br>19.65<br>23.58 | $2.9×10^4$ | 7 | $5.2×10^4$ |

A total of six experimental schemes were investigated: H5, H10, H20, and H50 under high cumulative electron dose conditions, and L5 and L50 under low cumulative electron dose conditions, where H and L denote the cumulative electron dose condition and the number indicates the laser pulse sequence employed. As an example, the measurement procedure of scheme H5 proceeded as follows. After an initial EELS spectrum measurement, five consecutive laser pulses were delivered to the GO film at the lowest fluence of 3.93 mJ/cm², followed by another core-loss EELS acquisition. The relative atomic percentages of carbon and oxygen were extracted from each spectrum to quantify the oxygen concentration. This cycle of 5 laser pulses followed by an EELS acquisition was repeated at the current fluence until the oxygen atomic percentage remained unchanged between consecutive acquisitions, indicating that the reduction rate had been exhausted and saturation had been reached. Only upon reaching saturation was the fluence incremented to 7.86 mJ/cm² followed by another set of laser irradiation and measurements until saturation was again reached. Then the fluence was increased and measurements made until saturation for 11.79 mJ/cm², 15.72 mJ/cm², and finally 19.65 mJ/cm², at which point the oxygen K-edge was no longer detectable. This continuous monitoring strategy resulted in a total of 43 EELS spectra collected throughout the experiment. **Figure 1(a)** shows the continuous increase in the cumulative electron dose and cumulative photon dose during the process of measurement scheme H5. For this scheme, the electron dose reached $3.2 \times 10^5$ e⁻/nm² by the end of the irradiation sequence.

In low electron dose schemes, such as L5, the sample was irradiated with same sequence of increasing laser fluences and number of laser pulses as that used in the associated high dose scheme (H5). However, EELS measurements were only made after completing the irradiation at a given fluence, before moving to the next laser fluence in the sequence. The GO film was therefore irradiated with the full predetermined number of pulses at each fluence step without interruption. The EELS spectrum acquired between fluence steps to was used to track the oxygen concentration through the process. This minimal EELS acquisition strategy for L5 is shown as the black line in **Figure 1(a)**, where the cumulative electron dose accumulates in discrete steps rather than continuously, reaching only $4.5 \times 10^4$ e⁻/nm² from just 6 EELS spectra nearly an order of magnitude lower than H5. Crucially, the cumulative photon dose delivered to the film remains identical between H5 and L5, illustrating that the two schemes differ exclusively in their cumulative electron dose.

Schemes H50 and L50 followed the same experimental approach as H5 and L5, respectively, with the key distinction that 50 consecutive laser pulses were delivered before each EELS acquisition rather than 5, and complete deoxygenation required an additional fluence step reaching a final value of 23.58 $mJ/cm^2$. As shown in **Figure 1(b)**, H50 yields a continuously increasing cumulative electron dose reaching $3.6 \times 10^5$ $e^-/nm^2$ from 49 EELS spectra, while L50 accumulates a cumulative electron dose of only $5.2 \times 10^4$ $e^-/nm^2$ from 7 spectra, again nearly an order of magnitude lower, with identical cumulative photon dose delivered in both cases. The colored regions in **Figure 1** demarcate the individual fluence intervals, illustrating how the cumulative electron dose and photon dose co-evolve continuously in the high electron dose schemes and discretely in the low electron dose schemes across each fluence step.

Importantly, across all schemes, $N\Phi_p$ remains identical between the high and low electron dose configurations for a given pulse sequence, as the same total number of laser pulses at equivalent fluence steps was delivered in each case. The key distinction between H and L schemes, therefore, lies exclusively in the frequency of EELS acquisition and consequently the cumulative electron dose deposited into the film, as clearly reflected in **Table 1** and **Figure 1**. While results from all schemes are relevant, the following discussion primarily focuses on the 5-pulse and 50-pulse sequences because they represent two contrasting reduction regimes. The 5-pulse sequence provides finer control over the cumulative photon dose, enabling a detailed examination of the stepwise evolution of GO during laser reduction, whereas the 50-pulse sequence represents a more aggressive laser irradiation strategy that facilitates comparison with higher cumulative fluence conditions. The intermediate H10 and H20 schemes are included in the supplementary material and verify the trends observed between these two limiting cases.

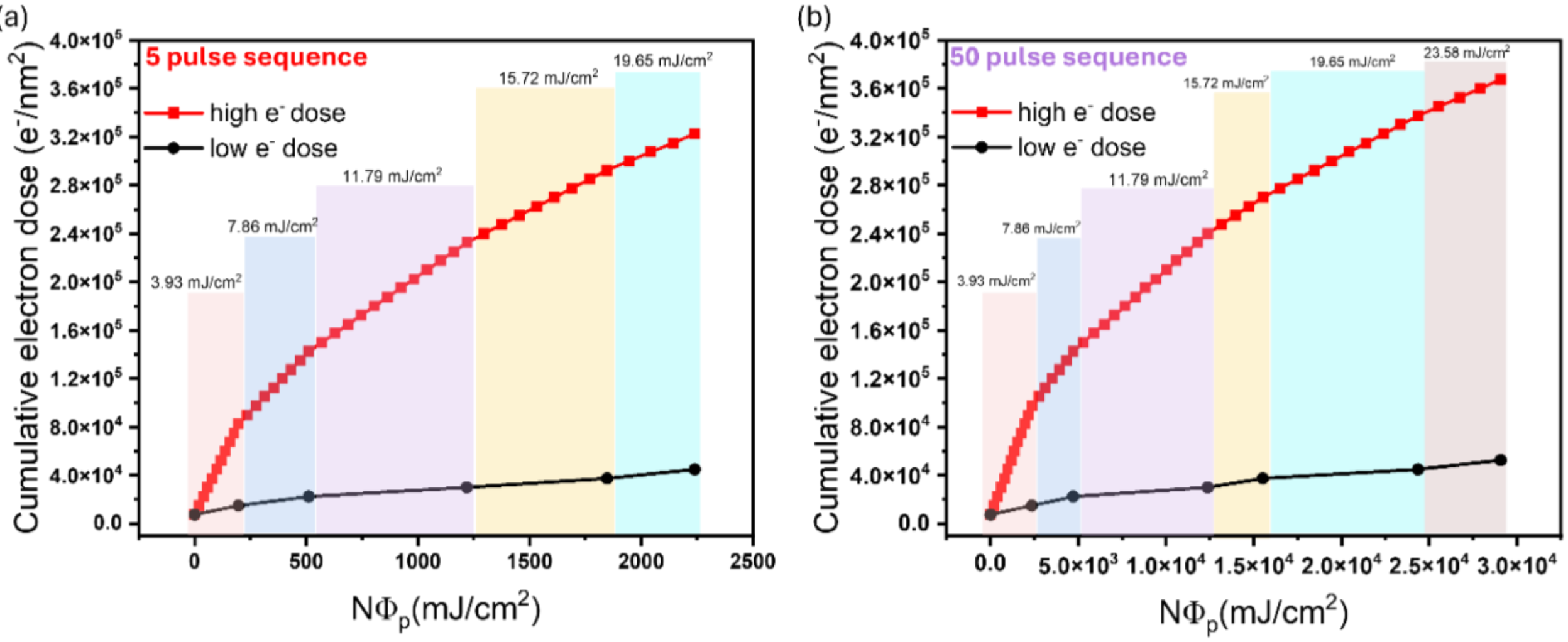

**Figure 1.** Cumulative electron dose as a function of cumulative photon dose ($N\Phi_p$) for the (a) 5-pulse and (b) 50-pulse sequences, shown for both high electron dose (red squares) and low electron dose (black circles) conditions. The colored regions denote the fluence intervals corresponding to each incremental step in the reduction process, with peak fluence values of 3.93, 7.86, 11.79, 15.72, and 19.65 mJ/cm$^2$ for the 5-pulse sequence, and an additional step at 23.58 mJ/cm$^2$ for the 50-pulse sequence irradiation schemes.

## 3. Results and Discussion

TEM micrographs and SAED patterns of the pristine GO sample prior to laser irradiation are shown in **Figure 2a**. The GO sheets are shown to be well distributed on the lacey carbon TEM grid, exhibiting uniform morphology with lateral regions of varying thickness. The local thickness, quantified using low-loss EELS following the log-ratio method [26], ranges from approximately 30 to 45 nm across the regions investigated. The corresponding SAED pattern displays diffuse rings characteristic of the largely disordered and turbostratic structure of pristine GO.

Following completion of stepwise laser irradiation under scheme H10, the GO sheets retained their continuous morphology after complete deoxygenation (**Figure 2b**), confirming that the gradual, controlled delivery of photon and electron dose prevented catastrophic structural damage. The SAED pattern acquired after reduction reveals a marked increase in diffraction ring intensity and the emergence of a discernible hexagonal pattern, indicating restoration of localized graphitic ordering and improved in-plane crystallinity associated with the conversion of GO to rGO [32].

To test the necessity of the stepwise approach a fresh GO region was irradiated with a single laser pulse at a fluence of 31.44 $mJ/cm^2$, corresponding to the highest value used in the scheme H10. Rather than producing a reduced film, this condition resulted in clear ablation, with GO sheets locally removed from the grid (**Figure 2c**). Therefore, direct high-fluence irradiation deposits energy at a rate exceeding the structural stability of the film and precludes any possibility of controlled reduction.

The chemical evolution of GO during laser-induced reduction following the scheme H10 experimental strategy was tracked directly using core-loss EELS, as shown in **Figure 2d**. The spectrum of pristine GO displays a well-defined C K-edge accompanied by a prominent O K-edge, confirming the abundance of oxygen functional groups across the lattice. Notably, the absence of the $\pi^*$ feature at ~285 eV in the pristine GO spectrum reflects the disruption of $sp^2$ bonding caused by the high density of these oxygen groups, which force carbon atoms into $sp^3$ configurations and break the conjugated network[33]. After stepwise laser irradiation, the EELS spectrum changes markedly. The O K-edge vanishes entirely, demonstrating complete removal of oxygen within the probed region **Figure 2d**. Simultaneously, a sharp $\pi^*$ feature emerges at ~285 eV, and the $\sigma^*$ peak intensity increases, together signaling the restoration of the in-plane $sp^2$ carbon network[24]. The correlated growth of both the $\pi^*$ and $\sigma^*$ features is particularly telling; it reflects not just partial conjugation, but a genuine reconstruction of the graphene-like lattice, where both the out-of-plane $\pi$ bonds and the in-plane C–C $\sigma$ framework are progressively recovered as oxygen is expelled[34].

This behavior was consistently observed across both the high electron dose and low electron dose approaches. In both cases, the development of $sp^2$ character tracked closely with the degree of oxygen removal; the more oxygen that left the system, the more pronounced the graphene-like spectral fingerprint became **Figure S1**. This is consistent with prior reports on thermally and chemically reduced GO, where the recovery of $sp^2$ bonding has been shown to scale with the extent of deoxygenation[34]. Taken together, these results confirm that stepwise increases in laser fluence irradiation is an effective strategy for achieving highly reduced graphene oxide, enabling controlled and progressive restoration of the conjugated carbon network.

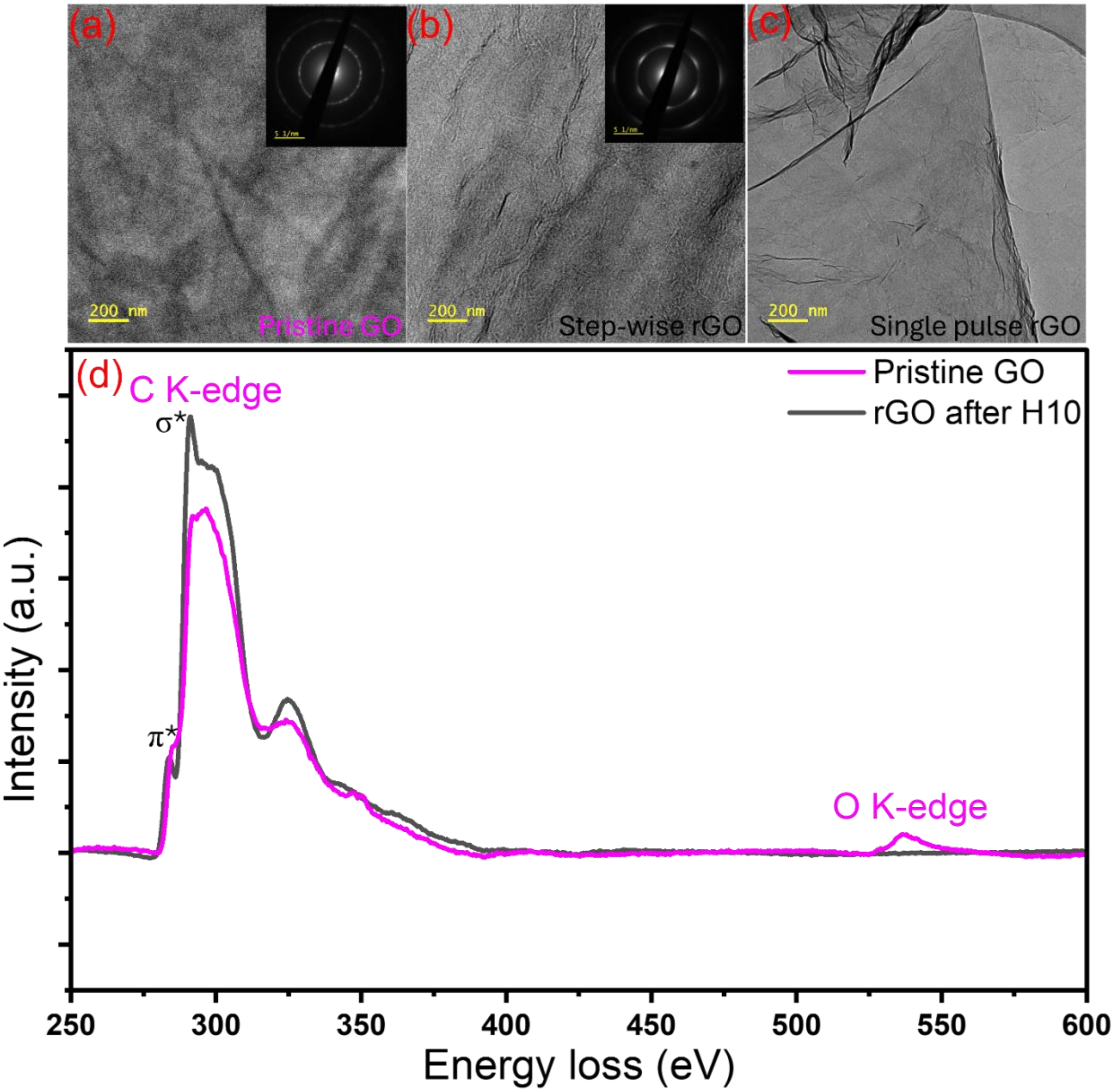

**Figure 2.** (a) TEM micrograph of pristine GO with the corresponding SAED pattern shown in the inset. (b) TEM micrograph of uniformly reduced GO obtained following scheme H10, with the inset displaying the corresponding SAED pattern. (c) TEM micrograph of rGO produced by a single laser pulse at a fluence of 31.44 mJ/cm$^2$. (d) EELS spectra before and after scheme H10 induced reduction.

To compare the extremes of the stepwise reduction strategy, the structural quality of films reduced under schemes H5, L5, H50, and L50 was evaluated by TEM imaging (**Figure 3a–d**). For schemes

H5 and L5, both high and low electron dose stepwise reductions produced continuous films without visible cracking after complete deoxygenation (**Figure 3a, b**), indicating that the 5-pulse sequence preserves the structural integrity of the film regardless of the cumulative electron dose. A more pronounced difference in film morphology is observed when comparing schemes H50 and L50. In scheme H50, the reduced films exhibit visible cracks and structural discontinuities (**Figure 3c**), indicating that the combination of high cumulative electron dose and the more aggressive 50-pulse sequence led to degradation of the graphitic network. In contrast, scheme L50 yielded structurally intact films with no visible cracking (**Figure 3d**), demonstrating that minimizing the cumulative electron dose during the 50-pulse reduction sequence is critical for preserving film continuity.

Taken together, these observations demonstrate that the cumulative electron dose plays a critical role in determining the structural integrity of the reduced films. Schemes H5, L5, and L50 all yield continuous, crack-free films upon complete deoxygenation, highlighting that both pulse sequences are capable of producing structurally intact reduced films when the cumulative electron dose is appropriately controlled. The only configuration resulting in visible cracking and structural discontinuities is scheme H50, suggesting that the high cumulative electron dose and high cumulative photon dose from the 50-pulse sequence per EELS spectra collectively exceeds the threshold for maintaining film integrity.

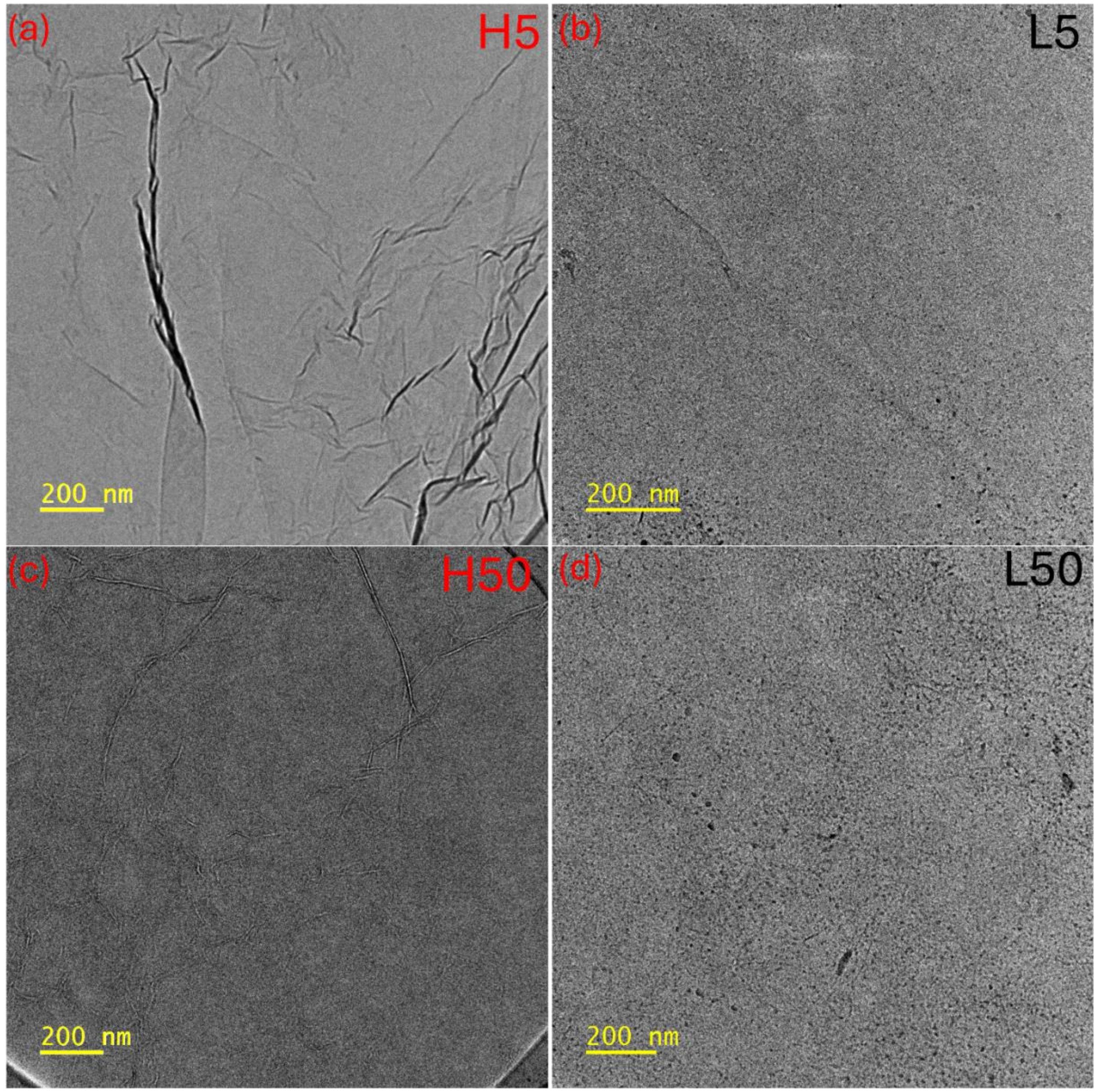

**Figure 3.** TEM micrographs of completely deoxygenated rGO films obtained under different irradiation schemes: (a) scheme H5, (b) scheme L5, (c) scheme H50, and (d) scheme L50.

The crystallographic quality of the reduced films was assessed by SAED and the corresponding radial intensity profiles for schemes H5, L5, H50, and L50, each compared against their respective pristine GO reference patterns (**Figure 4**). In the pristine GO, low-intensity diffraction rings corresponding to the (101) and (110) planes are observed, reflecting the largely disordered structure[7]. As the reduction proceeds, these rings become progressively more intense and better defined, indicating the growth of graphitic domains and improved localized in-plane structural ordering associated with restoration of $sp^2$ bonding [35]. The difference in diffraction ring positions between samples is attributed to variations in interlayer spacing, which is known to be influenced by both the degree of oxidation and the number of layers, as the d-spacing of GO has been shown to scale with oxygen content and film thickness [36].

For scheme H5, the SAED pattern shows an overall increase in Bragg reflection intensity relative to pristine GO (**Figure 4a, b**); however, the rings remain broad and continuous, indicating only limited development of hexagonal ordering despite complete oxygen removal. This is quantitatively confirmed by the radial intensity profile (**Figure 4c**), which reveals increased intensity at the (101) and (110) planes, consistent with partial restoration of the graphitic network as previously reported for thermally and optically reduced GO[36]. A weak (100) feature is visible at approximately 3.89 $nm^{-1}$, suggesting only nascent in-plane graphitic domain formation, with the higher cumulative electron dose partially suppressing localized $sp^2$ reconstruction [37-39].

In contrast, scheme L5 exhibits a markedly different structural response. The SAED pattern reveals well-defined discrete spots arranged along the rings (**Figure 4d, e**), consistent with restoration of hexagonal graphitic symmetry and improved localized ordering of $sp^2$ domains. The radial intensity profile (**Figure 4f**) further confirms a significantly greater increase in (101) and (110) intensity, alongside a well-resolved (100) reflection at approximately 3.89 $nm^{-1}$, corresponding to a d-spacing of 2.57 Å, broadly consistent with the in-plane reflection of graphene, with the slight deviation from the ideal value of 2.13 Å attributable to residual lattice strain or defects introduced during laser-driven reduction[36, 40]. Additionally, the (110) ring shifts from 8.7 $nm^{-1}$ to 8.5 $nm^{-1}$ following reduction in scheme L5, corresponding to an expansion of the in-plane lattice parameter consistent with defect formation during GO reduction[41, 42].

For scheme H50, the SAED pattern shows a notable decrease in diffraction intensity relative to pristine GO (**Figure 4g, h**), with weak diffuse rings reflecting degradation of the graphitic network consistent with the crack formation observed in TEM imaging. The radial intensity profile (**Figure 4i**) confirms only a modest increase in (101) and (110) intensity, attributed to the additional energy deposited into the film by the electron beam which, superimposed on the laser-induced thermal load, promotes mechanical failure and disrupts the restoration of long-range graphitic order.

Scheme L50, by contrast, displays the most pronounced structural improvement among all 50-pulse configurations. The SAED pattern exhibits strong, sharp Bragg reflections markedly exceeding those of pristine GO (**Figure 4j, k**), and the radial intensity profile (**Figure 4l**) confirms a significantly greater increase in (101) and (110) intensity relative to H50, indicative of extensive localized recovery of in-plane crystallographic order in the absence of beam-induced structural damage.

Taken together, across all four schemes, stepwise laser-induced reduction consistently improves in-plane graphitic order relative to pristine GO, with the highest degree of localized $sp^2$ reconstruction and crystallographic ordering achieved in scheme L5, highlighting that lowering the cummulative electron dose improves the ordering in the carbon network.

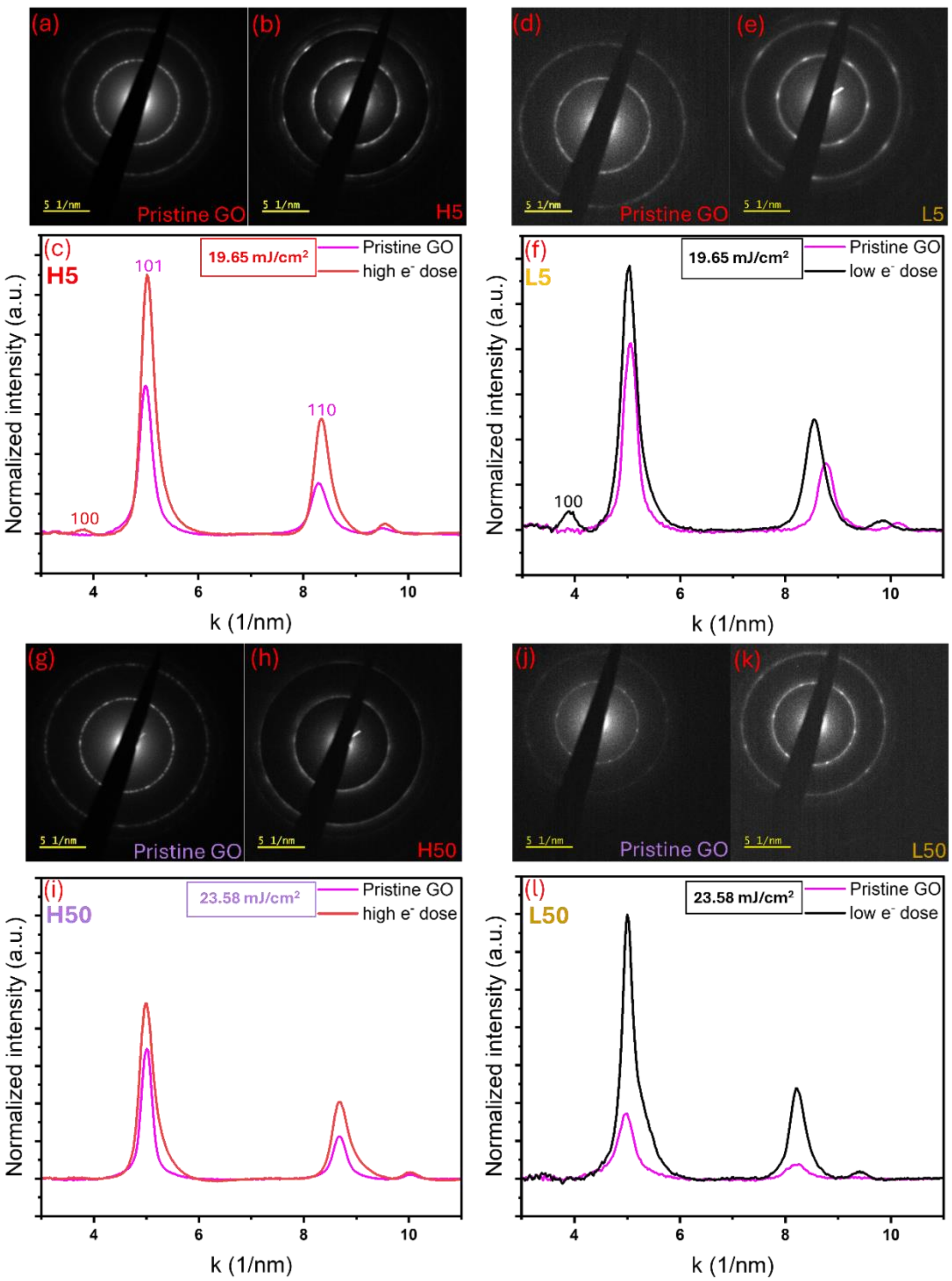

**Figure 4.** SAED patterns and corresponding radial intensity profiles of pristine GO and completely deoxygenated rGO for schemes H5, L5, H50, and L50. SAED patterns of pristine GO (a, d, g, j) and completely deoxygenated rGO (b, e, h, k) for schemes H5, L5, H50, and L50, respectively. Radial intensity profiles extracted from the corresponding SAED patterns plotted as a function of scattering vector, k ($nm^{-1}$), for schemes (c) H5, (f) L5, (i) H50, and (l) L50, comparing pristine GO (pink) with completely deoxygenated rGO (high electron dose in red, low electron dose in black). The (100), (101), and (110) in-plane reflections indicated where resolved. Complete deoxygenation was achieved at a peak fluence of 19.65 mJ/cm² for schemes H5 and L5, and 23.58 mJ/cm² for schemes H50 and L50.

Upon laser-induced reduction, GO films are expected to undergo structural compaction as oxygen functional groups and intercalated water molecules are removed from between the layers, leading to a collapse of the interlayer spacing and a reduction in overall film thickness[43]. To investigate this structural evolution across all four schemes, low-loss EELS spectra were acquired before and after complete reduction, and the relative film thickness was estimated using the log-ratio method. As shown in **Figure S2(a–d)**, the ZLP intensity increases following laser-induced reduction across all four schemes, consistent with a reduction in film thickness upon deoxygenation. However, a clear distinction is observed between the schemes. In scheme L5, the increase in ZLP intensity is notably smaller compared to schemes H5, H50, and L50, indicating significantly less film thinning despite achieving complete deoxygenation. This suggests that in scheme L5, the removal of oxygen functional groups proceeds in a more controlled manner, with the carbon lattice reorganizing gradually while maintaining the overall film architecture. In contrast, the more pronounced film thinning observed in schemes H5, H50, and L50 points to a more aggressive structural collapse, likely driven by the additional energy deposited by the higher cumulative electron or photon dose, accelerating interlayer collapse beyond what is induced by chemical deoxygenation alone. Taken together, the ZLP analysis confirms that scheme L5 achieves complete deoxygenation with the most controlled structural evolution, further validating it as the optimal configuration among all schemes investigated.

The carbon K-edge and oxygen K-edge EELS spectra acquired as a function of irradiation dose for schemes H5, L5, H50, and L50 provide direct spectroscopic evidence of the chemical and electronic structural changes induced by laser reduction across all experimental configurations **Figure 5(a, b)**. In all four cases, compared with pristine GO, the complete removal of the oxygen K-edge is observed, confirming successful deoxygenation regardless of the pulse number or electron dose. The clear recovery of the $\pi^*$ peak at ~285.0 eV, along with a pronounced increase in the σ* peak intensity across all spectra, is consistent with the successful restoration of the $sp^2$ graphitic network upon reduction, in agreement with previously reported carbon K-edge signatures for reduced graphene oxide[24]. The energy separation between the π* and σ* peaks is measured to be 7.0 eV and 7.5 eV for schemes H5/L5 and H50/L50, respectively, both in close agreement with the corresponding value reported for graphite (7.1 eV)[44], further validating the recovery of the localized graphitic carbon structure following laser-induced reduction.

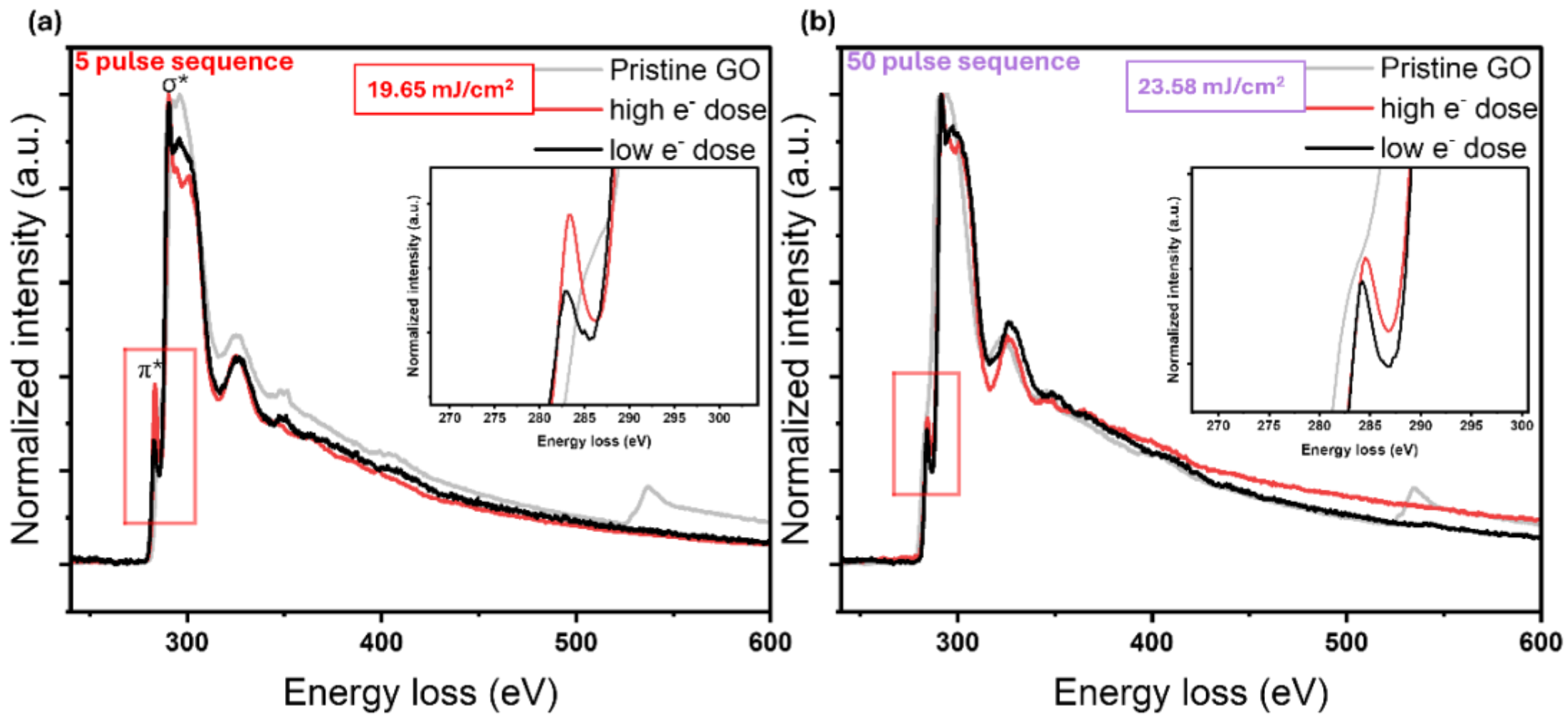

**Figure 5.** Carbon K-edge and oxygen K-edge EELS spectra of pristine GO and completely deoxygenated rGO for (a) schemes H5 and L5, and (b) schemes H50 and L50, acquired following complete deoxygenation at peak fluences of 19.65 mJ/cm$^2$ and 23.58 mJ/cm$^2$, respectively. The complete disappearance of the oxygen K-edge and the recovery of the characteristic π* and σ* features at the carbon K-edge confirm successful deoxygenation and restoration of the $sp^2$ graphitic network across all four schemes.

To elucidate the reduction kinetics and the role of cumulative electron dose, the oxygen atomic percentage obtained from quantitative analysis of the collected core-loss EELS spectra was plotted as a function of cumulative irradiated photon dose ($N\Phi_p$) for schemes H5, L5, H50, and L50 (**Figure 6a–d**). In all four schemes, the oxygen content decreases systematically with increasing cumulative photon dose, with each fluence step driving further deoxygenation until saturation is reached, confirming the effectiveness of the stepwise reduction strategy across both pulse sequences and electron dose conditions.

Comparing schemes H5 and L5 (**Figure 6a, b**), complete deoxygenation is achieved at a cumulative photon dose of approximately 2.5 × $10^3$ mJ/cm$^2$ in both cases, suggesting that the cumulative photon dose required for full oxygen removal is independent of the electron dose condition for the 5-pulse sequence. Similarly, schemes H50 and L50 (**Figure 6c, d**) both achieve complete deoxygenation at a cumulative photon dose of approximately 3 × $10^4$ mJ/cm$^2$, consistent with the higher cumulative photon dose required for the less efficient 50-pulse sequence. Notably, the cumulative electron dose differs by nearly an order of magnitude between the H and L schemes (**Table 1**). Yet complete deoxygenation is achieved in all cases, demonstrating that while the cumulative photon dose governs the overall deoxygenation threshold, the cumulative electron dose synergistically facilitates and accelerates the reduction process.

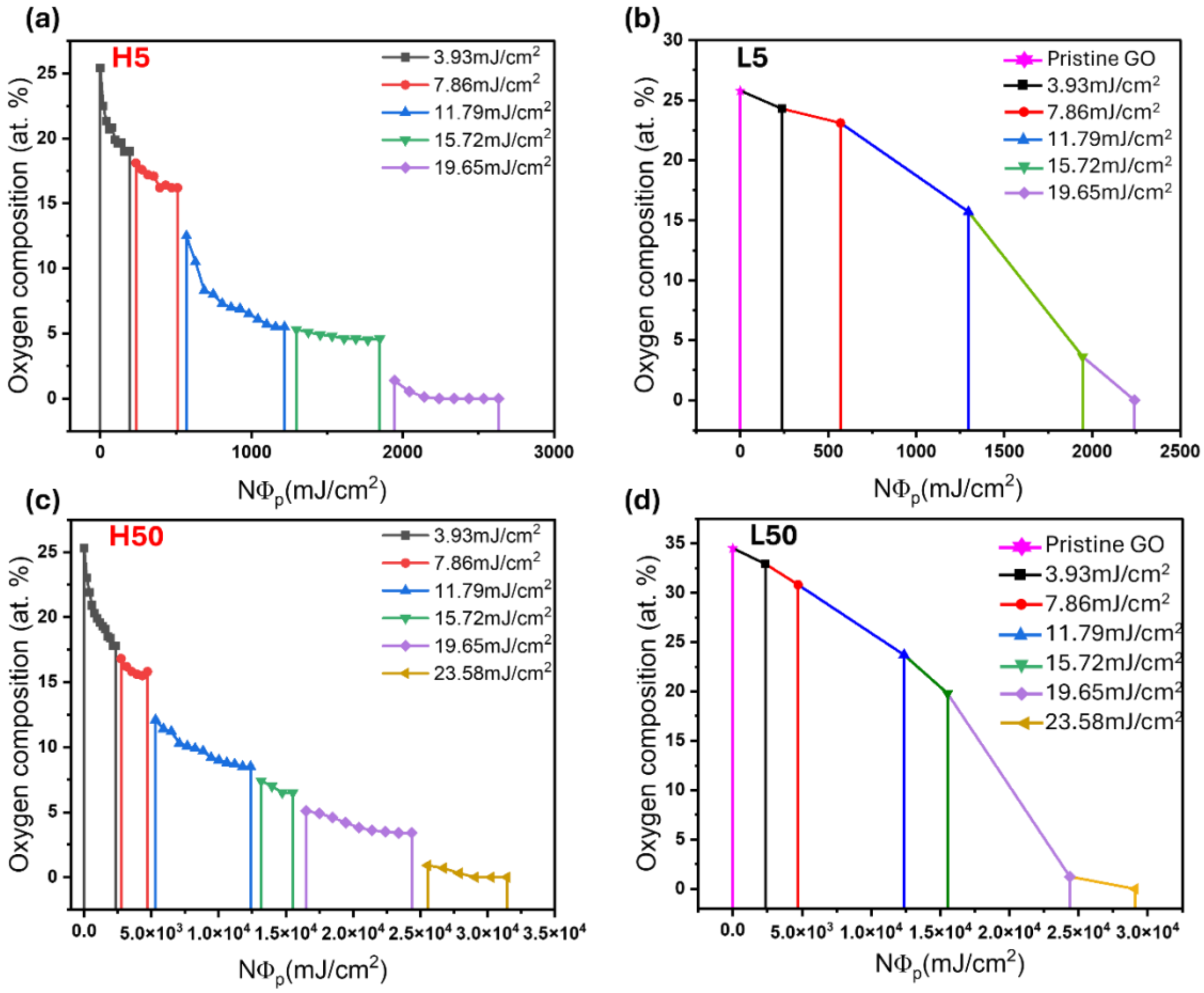

**Figure 6.** Evolution of oxygen composition (at. %) as a function of cumulative irradiated photon dose ($N\Phi_p$) for schemes (a) H5, (b) L5, (c) H50, and (d) L50, illustrating the different deoxygenation trends of GO under high and low electron dose conditions for the 5-pulse and 50-

pulse sequences, respectively. Each color represents a distinct fluence step, with vertical lines denoting the transition to the next fluence value upon saturation of oxygen removal at the preceding step.

A direct comparison of oxygen composition as a function of cumulative photon dose between the high and low electron dose conditions reveals the specific role of the electron beam in the reduction mechanism (**Figure 7a, b**). For both the 5-pulse and 50-pulse sequences, the total cumulative photon dose required to achieve complete deoxygenation remains essentially identical between schemes H5 and L5, and between H50 and L50, confirming that the final reduction outcome is governed primarily by the cumulative photon dose rather than the cumulative electron dose.
However, a clear and reproducible difference in the initial rate of oxygen removal is observed. In schemes H5 and H50, the oxygen content decreases more rapidly at low cumulative photon doses compared to their low electron dose counterparts L5 and L50, indicating that the electron beam accelerates the early stages of the reduction process. This electron-beam-assisted effect accounts for approximately the first ~5 at.% reduction in oxygen content **Figure S3**, beyond which the two curves converge, and the photon dose becomes the dominant driver of deoxyg enation, bringing the reduction to completion regardless of the cumulative electron dose history. This behavior is consistent with radiolytic effects and vacancy creation, reported in the electron beam modification of carbon-based materials [45, 46], wherein repeated electron beam exposure activates or destabilizes oxygen-containing functional groups in the GO lattice[47], effectively lowering the energy barrier for subsequent laser-driven deoxygenation. Once this initial activation threshold is surpassed, the electron beam-induced modification of the GO lattice renders the material more responsive to subsequent light irradiation, effectively lowering the energy barrier for laser-driven deoxygenation. As a result, the stepwise photon dose takes over as the principal reduction mechanism, driving complete deoxygenation independently of the electron dose condition.

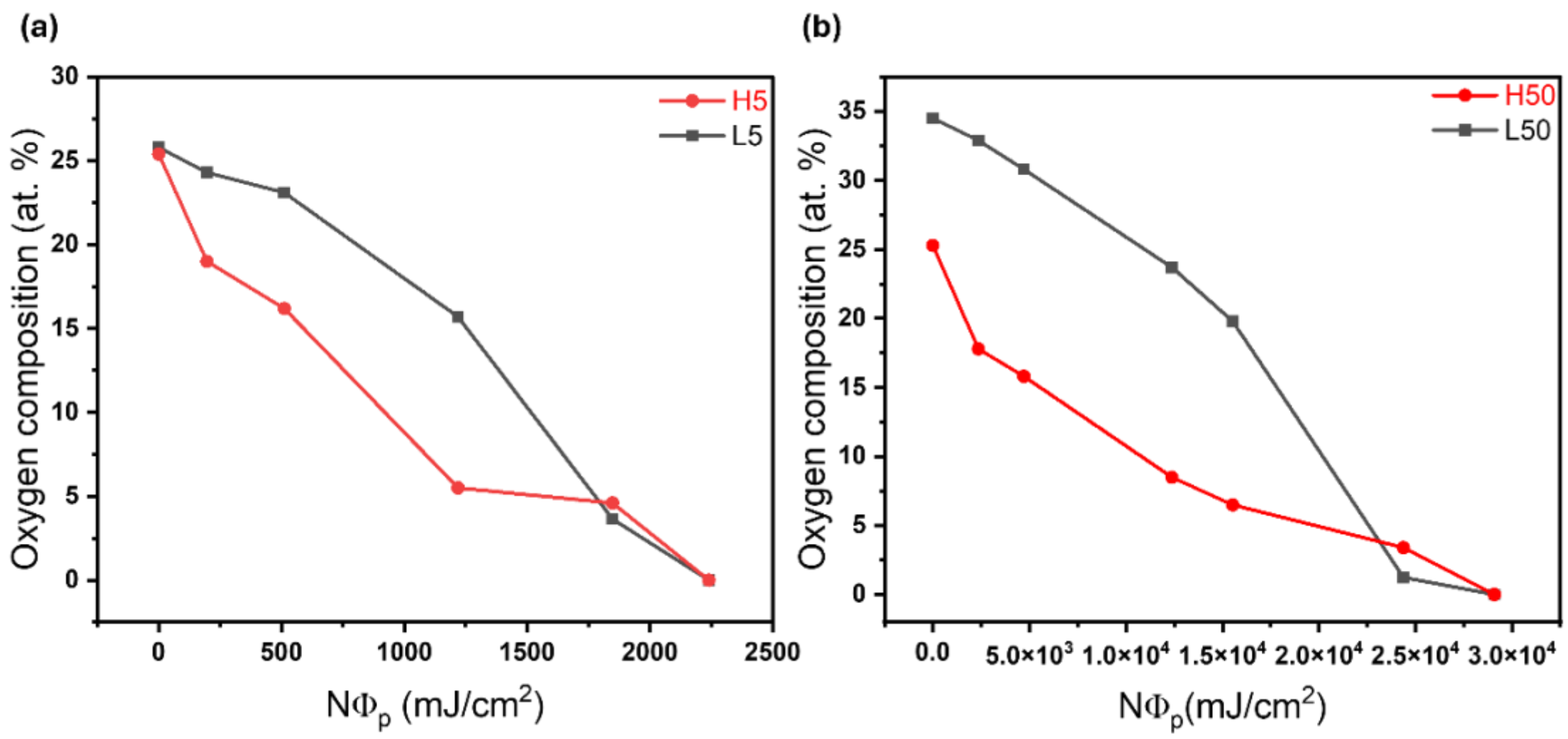

**Figure 7.** Direct comparison of oxygen composition (at.%) as a function of cumulative photon dose ($N\Phi_p$) between high (H, red circles) and low (L, black squares) electron dose conditions for (a) the 5-pulse sequences (schemes H5 and L5) and (b) the 50-pulse sequences (schemes H50 and L50).

Finally, to further elucidate the dependence of reduction efficiency on pulse sequence, the evolution of oxygen composition as a function of cumulative photon dose is shown for schemes H10 and H20 in **Figure S4(a, b)**. Consistent with the trend established for schemes H5 and H50, both H10 and H20 achieve complete deoxygenation through the same stepwise fluence increments. However, a clear and systematic increase in the required cumulative photon dose with increasing pulse sequence is observed from approximately $2.5 \times 10^3$ mJ/cm² for H5, to $4.3 \times 10^3$ mJ/cm² for H10, $1.4 \times 10^4$ mJ/cm² for H20, and $2.9 \times 10^4$ mJ/cm² for H50.

Since all pulse sequences share the same repetition rate of 10 Hz, with 0.1 seconds between consecutive pulses far exceeding the thermal cooling timescale of the film thermal accumulation between pulses can be entirely excluded as a contributing factor [46]. Instead, the observed dependence of cumulative photon dose on pulse sequence is attributed to the difference in frequency of electron beam intervention between pulse sequences. In lower pulse sequences such as H5, EELS acquisition and thus electron beam exposure occurs more frequently per unit of cumulative photon dose delivered, continuously generating defects that enhance light absorption efficiency and facilitate oxygen removal with each subsequent laser pulse, as previously described. In higher pulse sequences such as H50, electron beam intervention occurs less frequently per unit of photon dose, resulting in slower accumulation of beam-induced defects and consequently a higher cumulative photon dose required to achieve complete deoxygenation. This interpretation is consistent across all four high electron dose schemes and is further supported by the established role of electron beam irradiation in directly promoting deoxygenation of GO through vacancy creation and oxygen removal [48, 49].

## 4. Conclusions

Our results show that the coupled electron beam and stepwise laser fluence irradiation of graphene oxide, when carefully optimized in terms of pulse sequence, cumulative photon dose, and cumulative electron dose, achieves what has proven difficult - complete deoxygenation while preserving the structural integrity of the film. Among all experimental schemes investigated, L5 emerges as the optimal configuration, yielding highly reduced graphene oxide with superior localized in-plane crystallographic order, evidenced by the appearance of the (100) reflection and enhanced intensity of the (101) and (110) planes in electron diffraction, alongside full recovery of the $sp^2$ graphitic network confirmed by a $\pi^*-\sigma^*$ energy separation of 7.0 eV, in close agreement with that of graphite (7.1 eV).

We propose that the cumulative electron dose is not a passive monitoring parameter but an active contributor to the reduction mechanism. The electron beam accounts for approximately 5 at. % of the initial oxygen removal and improves the efficiency of the process by increasing the reduction rate for subsequent laser irradiation. Beyond this initial contribution, the cumulative photon dose governs the overall deoxygenation threshold, with lower pulse sequences achieving complete oxygen removal at significantly reduced cumulative photon doses owing to more frequent electron beam intervention per unit of delivered photon dose. Conversely, excessive cumulative electron dose, as demonstrated in scheme H50, induces crack formation and disrupts $sp^2$ network recovery, underscoring the critical balance that must be maintained between photon and electron dose for optimal structural outcomes. The stepwise laser fluence irradiation strategy introduced here, with its inherent spatial selectivity, ambient processing compatibility, and precise tunability can be extended and suggests a viable pathway toward the fabrication of large-area, high-quality, highly-reduced rGO films.

## References

[1] L.G. Guex, B. Sacchi, K.F. Peuvot, R.L. Andersson, A.M. Pourrahimi, V. Ström, S. Farris, R.T. Olsson, Experimental review: chemical reduction of graphene oxide (GO) to reduced graphene oxide (rGO) by aqueous chemistry, Nanoscale 9(27) (2017) 9562-9571.

[2] G. Eda, M. Chhowalla, Chemically derived graphene oxide: towards large-area thin-film electronics and optoelectronics, Advanced materials 22(22) (2010) 2392-2415.
[3] A. Betancur, N. Ornelas-Soto, A. Garay-Tapia, F. Pérez, Á. Salazar, A.G. García, A general strategy for direct synthesis of reduced graphene oxide by chemical exfoliation of graphite, Materials Chemistry and Physics 218 (2018) 51-61.
[4] L. Zhang, J. Liang, Y. Huang, Y. Ma, Y. Wang, Y. Chen, Size-controlled synthesis of graphene oxide sheets on a large scale using chemical exfoliation, Carbon 47(14) (2009) 3365-3368.
[5] C.V. Gomez, E. Robalino, D. Haro, T. Tene, P. Escudero, A. Haro, J. Orbe, Structural and electronic properties of graphene oxide for different degree of oxidation, Materials Today: Proceedings 3(3) (2016) 796-802.
[6] N.S. Suhaimin, M.F.R. Hanifah, M. Azhar, J. Jaafar, M. Aziz, A. Ismail, M. Othman, M.A. Rahman, F. Aziz, N. Yusof, The evolution of oxygen-functional groups of graphene oxide as a function of oxidation degree, Materials Chemistry and Physics 278 (2022) 125629.
[7] K. Krishnamoorthy, M. Veerapandian, K. Yun, S.-J. Kim, The chemical and structural analysis of graphene oxide with different degrees of oxidation, Carbon 53 (2013) 38-49.
[8] S. Pei, H.-M. Cheng, The reduction of graphene oxide, Carbon 50(9) (2012) 3210-3228.
[9] S.D. Perera, R.G. Mariano, N. Nijem, Y. Chabal, J.P. Ferraris, K.J. Balkus Jr, Alkaline deoxygenated graphene oxide for supercapacitor applications: An effective green alternative for chemically reduced graphene, Journal of Power Sources 215 (2012) 1-10.
[10] K. De Silva, H.-H. Huang, R. Joshi, M. Yoshimura, Chemical reduction of graphene oxide using green reductants, Carbon 119 (2017) 190-199.
[11] I.K. Moon, J. Lee, R.S. Ruoff, H. Lee, Reduced graphene oxide by chemical graphitization, Nature communications 1(1) (2010) 73.
[12] M.A.A. Farid, J. Lease, Y. Andou, Temperature-dependent deoxygenation during sulfonation for graphene oxide reduction, Applied Surface Science 691 (2025) 162678.
[13] J. Cao, G.-Q. Qi, K. Ke, Y. Luo, W. Yang, B.-H. Xie, M.-B. Yang, Effect of temperature and time on the exfoliation and de-oxygenation of graphite oxide by thermal reduction, Journal of Materials Science 47(13) (2012) 5097-5105.
[14] G.V. Murastov, A.A.e. Lipovka, M.I.g. Fatkullin, R.D. Rodriguez, E.S. Sheremet, Laser reduction of graphene oxide: local control of material properties, Physics-Uspekhi 66(11) (2023) 1105-1133.
[15] Y. Zhou, Q. Bao, B. Varghese, L.A.L. Tang, C.K. Tan, C.H. Sow, K.P. Loh, Microstructuring of graphene oxide nanosheets using direct laser writing, Advanced Materials 22(1) (2010) 67-71.
[16] M.J. Low, H. Lee, C.H.J. Lim, C.S. Sandeep, V.M. Murukeshan, S.-W. Kim, Y.-J. KIm, Laser-induced reduced-graphene-oxide micro-optics patterned by femtosecond laser direct writing, Applied Surface Science 526 (2020) 146647.
[17] M. Kasischke, S. Maragkaki, S. Volz, A. Ostendorf, E.L. Gurevich, Simultaneous nanopatterning and reduction of graphene oxide by femtosecond laser pulses, Applied Surface Science 445 (2018) 197-203.
[18] Y.L. Zhang, L. Guo, H. Xia, Q.D. Chen, J. Feng, H.B. Sun, Photoreduction of graphene oxides: methods, properties, and applications, Advanced Optical Materials 2(1) (2014) 10-28.
[19] C.-R. Yang, S.-F. Tseng, Y.-T. Chen, Laser-induced reduction of graphene oxide powders by high pulsed ultraviolet laser irradiations, Applied Surface Science 444 (2018) 578-583.
[20] B. De Lima, M.I.B. Bernardi, V.R. Mastelaro, Wavelength effect of ns-pulsed radiation on the reduction of graphene oxide, Applied Surface Science 506 (2020) 144808.
[21] V. Agarwal, P.B. Zetterlund, Strategies for reduction of graphene oxide–A comprehensive review, Chemical Engineering Journal 405 (2021) 127018.
[22] W. Yu, L. Sisi, Y. Haiyan, L. Jie, Progress in the functional modification of graphene/graphene oxide: A review, RSC advances 10(26) (2020) 15328-15345.
[23] I. Fatimah, G. Fadillah, R.A. Rednasari, S. Wahyuningsih, Green reduction of graphene oxide using Annona muricata leaves extract for adsorption of methylene blue, Inorganic Chemistry Communications 146 (2022) 110144.
[24] A. Tararan, A. Zobelli, A.M. Benito, W.K. Maser, O. Stéphan, Revisiting graphene oxide chemistry via spatially-resolved electron energy loss spectroscopy, Chemistry of Materials 28(11) (2016) 3741-3748.
[25] M. Pelaez-Fernandez, A. Bermejo, A.M. Benito, W.K. Maser, R. Arenal, Detailed thermal reduction analyses of graphene oxide via in-situ TEM/EELS studies, Carbon 178 (2021) 477-487.
[26] I. Ali, H. Lei, S. Sun, K.R. Beyerlein, Detailed in situ TEM/EELS analysis of laser-induced reduction of graphene oxide, Carbon (2025) 120444.

[27] T. Malis, S. Cheng, R. Egerton, EELS log‐ratio technique for specimen‐thickness measurement in the TEM, Journal of electron microscopy technique 8(2) (1988) 193-200.
[28] H. Shinotsuka, S. Tanuma, C.J. Powell, D.R. Penn, Calculations of electron inelastic mean free paths. X. Data for 41 elemental solids over the 50 eV to 200 keV range with the relativistic full Penn algorithm, Surface and Interface Analysis 47(9) (2015) 871-888.
[29] K. Iakoubovskii, K. Mitsuishi, Y. Nakayama, K. Furuya, Mean free path of inelastic electron scattering in elemental solids and oxides using transmission electron microscopy: Atomic number dependent oscillatory behavior, Physical Review B 77(10) (2008) 104102.
[30] W. Ding, T. Chen, K. Liao, L. He, F. Song, J. Zhou, J. Wan, G. Wang, M. Han, Scaling the dynamic electron scattering in imaging the graphene sheets by the high-angle annular dark-field microscopy, Journal of Nanoscience and Nanotechnology 12(8) (2012) 6494-6498.
[31] F. Song, Z. Li, Z. Wang, L. He, M. Han, G. Wang, Free-standing graphene by scanning transmission electron microscopy, Ultramicroscopy 110(12) (2010) 1460-1464.
[32] H.-J. Lee, J.S. Kim, K.Y. Lee, K.H. Park, J.-S. Bae, M. Mubarak, H. Lee, Elucidation of an intrinsic parameter for evaluating the electrical quality of graphene flakes, Scientific Reports 9(1) (2019) 557.
[33] B.S. Lee, Y. Lee, J.Y. Hwang, Y.C. Choi, Structural properties of reduced graphene oxides prepared using various reducing agents, Carbon letters 16(4) (2015) 255-259.
[34] C.-H. Chuang, Y.-F. Wang, Y.-C. Shao, Y.-C. Yeh, D.-Y. Wang, C.-W. Chen, J. Chiou, S.C. Ray, W. Pong, L. Zhang, The effect of thermal reduction on the photoluminescence and electronic structures of graphene oxides, Scientific reports 4(1) (2014) 4525.
[35] S. Gupta, P. Joshi, J. Narayan, Electron mobility modulation in graphene oxide by controlling carbon melt lifetime, Carbon 170 (2020) 327-337.
[36] M. Ceniceros-Reyes, K. Marín-Hernández, U. Sierra, E. Saucedo-Salazar, R. Mendoza-Resendez, C. Luna, P. Hernández-Belmares, O. Rodríguez-Fernández, S. Fernández-Tavizón, E. Hernández-Hernández, Reduction of graphene oxide by in-situ heating experiments in the transmission electron microscope, Surfaces and Interfaces 35 (2022) 102448.
[37] T. Latychevskaia, W.-H. Hsu, W.-T. Chang, C.-Y. Lin, I.-S. Hwang, Three-dimensional surface topography of graphene by divergent beam electron diffraction, Nature communications 8(1) (2017) 14440.
[38] W. Zhang, W.-C. Lu, H.-X. Zhang, K. Ho, C. Wang, Lattice distortion and electron charge redistribution induced by defects in graphene, Carbon 110 (2016) 330-335.
[39] D. Manno, L. Torrisi, L. Silipigni, A. Buccolieri, M. Cutroneo, A. Torrisi, L. Calcagnile, A. Serra, From GO to rGO: an analysis of the progressive rippling induced by energetic ion irradiation, Applied Surface Science 586 (2022) 152789.
[40] T.T. Tran, P.O. Persson, N. Pham, R. Holenak, D. Primetzhofer, Mobility of single vacancies and adatoms in graphene at room temperature, Small 21(35) (2025) 2504370.
[41] C. Gómez-Navarro, J.C. Meyer, R.S. Sundaram, A. Chuvilin, S. Kurasch, M. Burghard, K. Kern, U. Kaiser, Atomic structure of reduced graphene oxide, Nano Lett. 10(4) (2010) 1144-1148.
[42] T. Taniguchi, L. Nurdiwijayanto, N. Sakai, K. Tsukagoshi, T. Sasaki, T. Tsugawa, M. Koinuma, K. Hatakeyama, S. Ida, Revisiting the two-dimensional structure and reduction process of graphene oxide with in-plane X-ray diffraction, Carbon 202 (2023) 26-35.
[43] Z. Wan, S. Wang, B. Haylock, J. Kaur, P. Tanner, D. Thiel, R. Sang, I.S. Cole, X. Li, M. Lobino, Tuning the sub-processes in laser reduction of graphene oxide by adjusting the power and scanning speed of laser, Carbon 141 (2019) 83-91.
[44] K.A. Mkhoyan, A.W. Contryman, J. Silcox, D.A. Stewart, G. Eda, C. Mattevi, S. Miller, M. Chhowalla, Atomic and electronic structure of graphene-oxide, Nano Lett. 9(3) (2009) 1058-1063.
[45] F. Banhart, The Formation and Transformation of Low‐Dimensional Carbon Nanomaterials by Electron Irradiation, Small 21(28) (2025) 2310462.
[46] I. Ali, H. Mba, M. Picher, S. Verma, F. Banhart, K.R. Beyerlein, Fast reduction of electron-beam-activated graphene oxide by an infrared laser pulse, arXiv preprint arXiv:2605.01125 (2026).
[47] N.S.B.M. Nasir, F.P. Chee, N. Johrin, P.Y. Moh, M.M.I.B.M. Hasnan, M. Al-Buriahi, J.H.W. Chang, S. Salleh, S. Ibrahim, H. Hasim, Structural and chemical modifications of graphene oxide induced by high-energy electron beam irradiation, Journal of Materials Science: Materials in Electronics 37(5) (2026) 377.
[48] Y. Yang, L. Chen, D.-Y. Li, R.-B. Yi, J.-W. Mo, M.-H. Wu, G. Xu, Controllable reduction of graphene oxide by electron-beam irradiation, RSC advances 9(7) (2019) 3597-3604.

[49] C. Tyagi, G. Lakshmi, S. Kumar, A. Tripathi, D. Avasthi, Structural changes in graphene oxide thin film by electron-beam irradiation, Nuclear Instruments and Methods in Physics Research Section B: Beam Interactions with Materials and Atoms 379 (2016) 171-175.

**Stepwise laser-induced Complete Deoxygenation and Enhanced Localized $sp^2$ Character of Graphene Oxide Revealed by in-situ TEM**

Israt Ali[1], Kenneth R. Beyerlein[1]*

[1]Institut National de la Recherche Scientifique (INRS), Center Énergie Matériaux Télécommunications, Varennes, Québec, J3X 1P7, Canada

*Corresponding Author: kenneth.beyerlein@inrs.ca

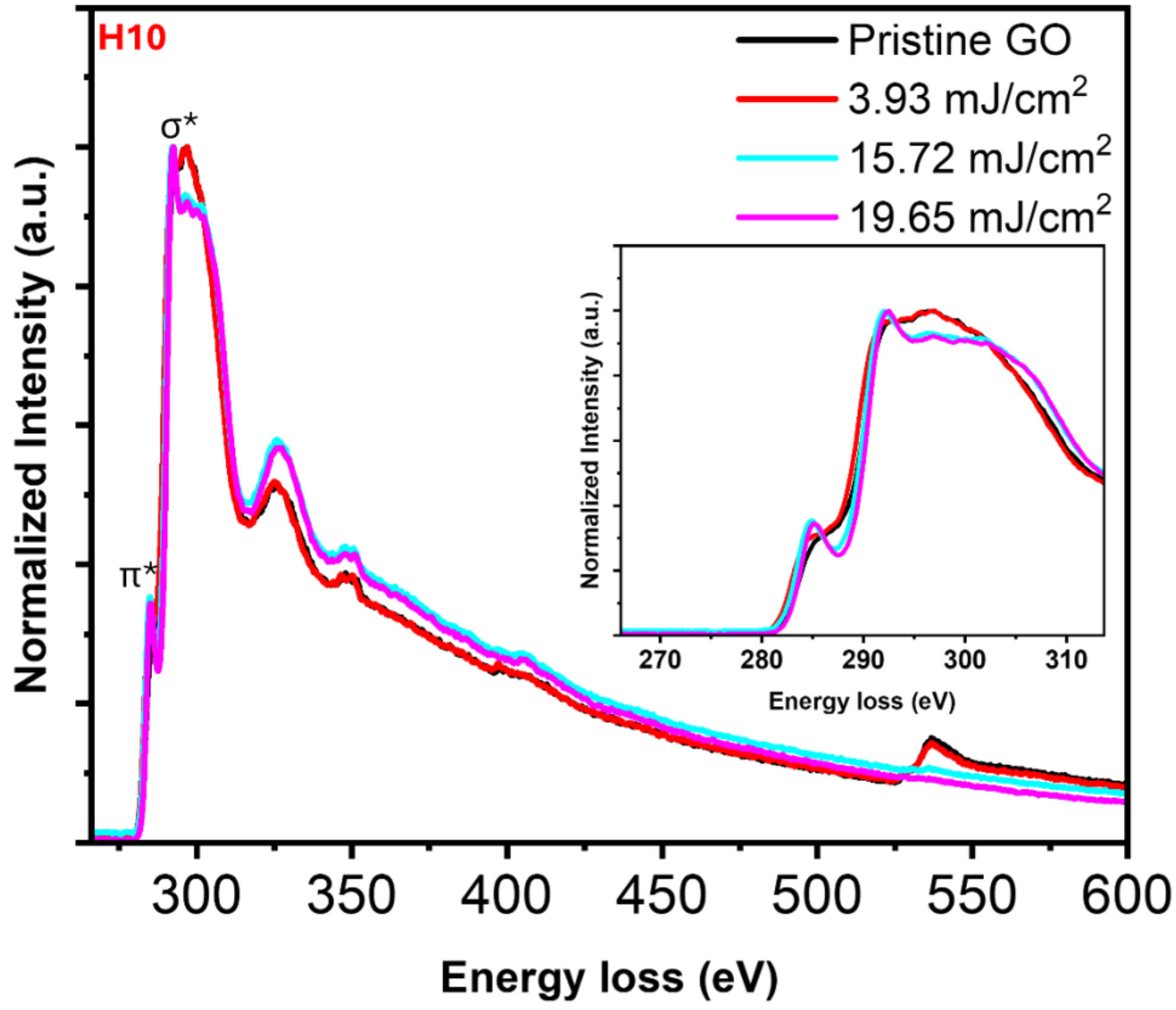

Figure S1. Carbon K-edge and oxygen K-edge EELS spectra acquired during stepwise laser-induced reduction under scheme H10, shown at selected cumulative fluences of 3.93, 15.72, and 19.65 mJ/cm², alongside the pristine GO reference spectrum. The inset displays the magnified carbon K-edge region (265–315 eV), highlighting the progressive evolution of the π* and σ* features with increasing cumulative fluence. The gradual suppression of the oxygen K-edge at ~532 eV and the concurrent sharpening of the π* peak at 285.0 eV confirm the stepwise removal of oxygen functional groups and progressive restoration of the sp² graphitic network with increasing cumulative photon dose.

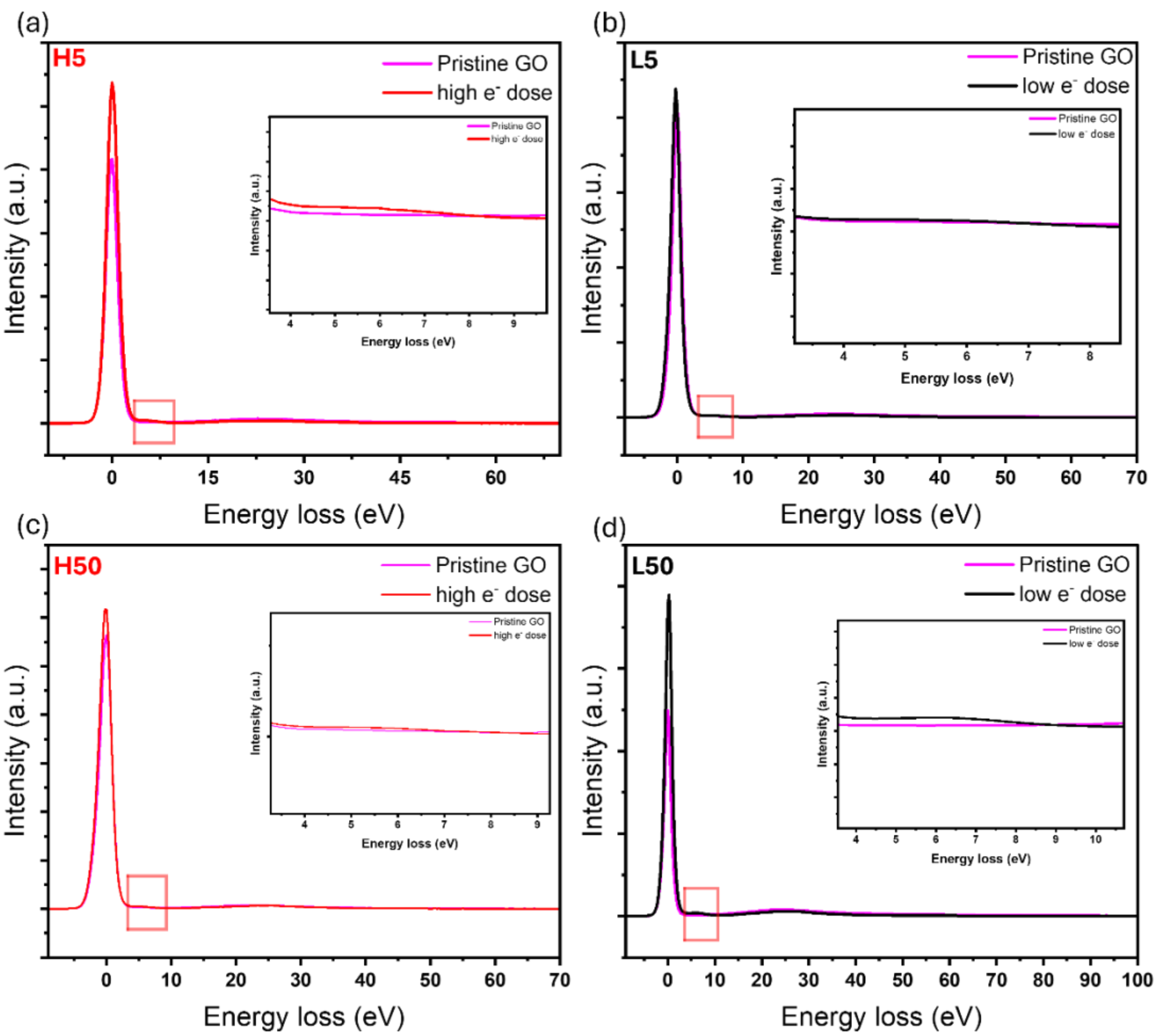

Figure S. Low-loss EELS spectra showing the zero-loss peak (ZLP) of pristine GO and final rGO for schemes (a) H5, (b) L5, (c) H50, and (d) L50, acquired at the same sample region before and after complete laser-induced reduction. The insets show the magnified low-energy region (3–10 eV), highlighting the change in ZLP intensity upon reduction.

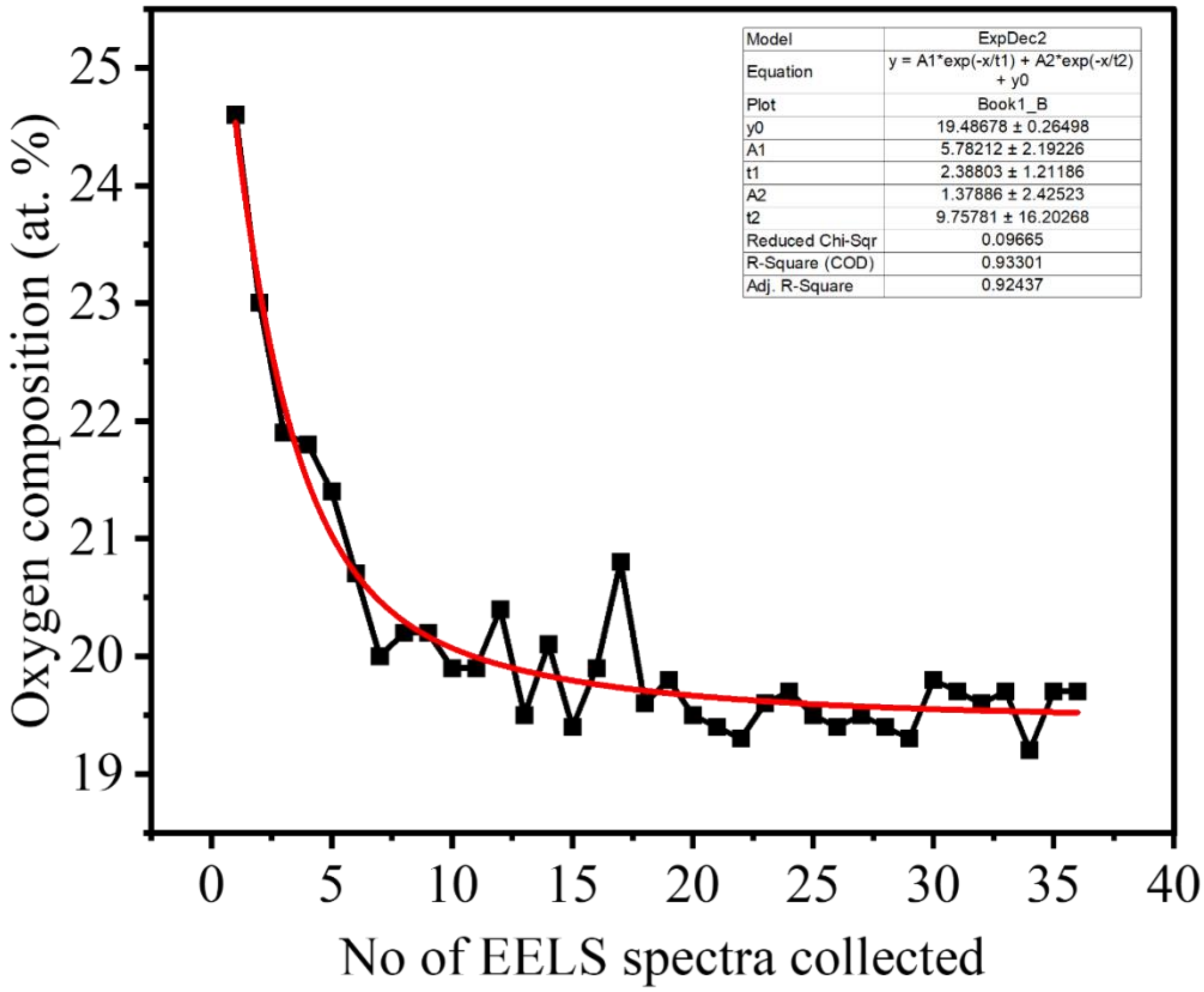

Figure S3. Evolution of oxygen composition (at.%) as a function of the number of EELS spectra collected during electron beam irradiation of pristine GO without laser irradiation, demonstrating the standalone contribution of the electron beam to deoxygenation. The data are fitted with a double exponential decay function (red curve, $R^2 = 0.93$), revealing a rapid initial reduction in oxygen content within the first ~5 EELS acquisitions, followed by a gradual plateau at approximately 19.5 at.%, beyond which no further oxygen removal is observed. This confirms that the electron beam alone is capable of driving only partial deoxygenation of GO, accounting for approximately 5 at.% reduction in oxygen content, and that complete deoxygenation requires the synergistic contribution of cumulative photon dose from laser irradiation.

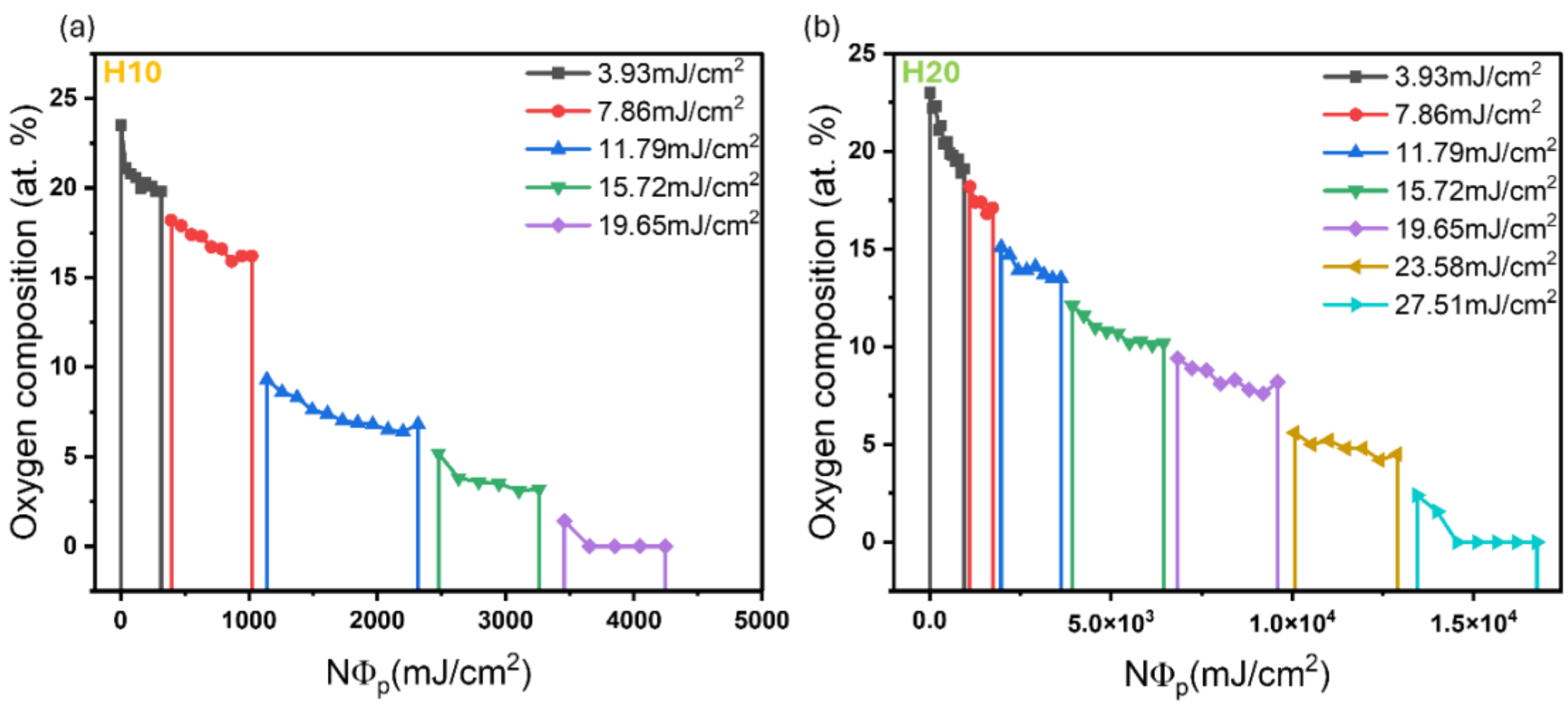

Figure S4. Evolution of oxygen composition (at.%) as a function of cumulative photon dose ($N\Phi_p$) for schemes (a) H10 and (b) H20 under high electron dose conditions. Each color represents a distinct fluence step, with vertical lines denoting the transition to the next fluence value upon saturation of oxygen removal at the preceding step. Complete deoxygenation is achieved at cumulative photon doses of approximately $4.3 \times 10^3$ mJ/cm² and $1.4 \times 10^4$ mJ/cm² for H10 and H20, respectively. This trend is consistent with the systematic increase in required cumulative photon dose with increasing pulse sequence observed across all high electron dose schemes (H5, H10, H20, H50).